\documentclass[structabstract]{aa}
\usepackage{txfonts}
\usepackage{graphicx}
\usepackage{natbib}

\bibpunct{(}{)}{;}{a}{}{,} % to follow the A&A style

\newcommand{\dx}{\mathrm{d}}

\def\phxoned{{\texttt{PHOENIX/1D}}}

\begin{document}

\title{A 3D radiative transfer framework}
\subtitle{IX. Time dependence}
 \author{D. Jack\inst{\ref{inst1},\ref{inst2}}
  \and P. H. Hauschildt\inst{\ref{inst1}}
   \and E. Baron\inst{\ref{inst1},\ref{inst3}}}

\institute{Hamburger Sternwarte, Gojenbergsweg 112, 21029 Hamburg, Germany\\
\email{djack@hs.uni-hamburg.de; yeti@hs.uni-hamburg.de}\label{inst1}
  \and
Departamento de Astronom\'\i{}a, Universidad de Guanajuato, Apartado Postal 144, 36000, Guanajuato, Mexico\label{inst2}
  \and
Homer L. Dodge Department of Physics and Astronomy, University of Oklahoma, 440 W Brooks, Rm 100, Norman, OK 73019-2061 USA\\
\email{baron@ou.edu}\label{inst3}}

\date{Received 26 September 2011 /
      Accepted 29 August 2012}

\abstract {Time-dependent, 3D  radiation transfer calculations are important for the modeling of
a variety of objects, from supernovae and novae to simulations of stellar variability and  activity.
Furthermore, time-dependent calculations can be used to obtain a 3D radiative equilibrium model structure via
relaxation in time.}
{We extend our 3D radiative transfer framework to include direct time dependence of
the radiation field; i.e., the $\partial I/\partial t$ terms are fully considered
in the solution of radiative transfer problems.}
{We build on the framework  that we have described in previous papers in this series and develop
a subvoxel method for the $\partial I/\partial t$  terms.}
{We test the implementation by comparing the 3D results to our well
tested 1D time dependent radiative transfer code in spherical symmetry.
A simple 3D test model is also presented.}
{The 3D time dependent radiative transfer method is now included in our 3D RT framework and
in PHOENIX/3D.}


\maketitle

\section{Introduction}

Supernovae of all types undergo a rather rapid evolution after their explosion. During the free-expansion phase, observations 
show fast evolving light curves and changing spectral features.
In type Ia supernovae, the radioactive decay of $^{56}$Ni heats the envelope,
causing a peak in the optical light curve about 20 days after the explosion.
We have already modeled light curves of SNe Ia with our 1D spherically symmetric model atmosphere code \phxoned\ \citep{jack11,jack12,wang12}.
To compute more accurate model light curves, the hydrodynamical evolution during the free expansion phase needs
to be calculated in 3D, including the time dependence of the radiation field.
This requires a time-dependent solution of the 3D radiative transfer equation, including the velocity field.
In addition, even for the calculation of static and stationary 3D atmospheres, time relaxation to radiative equilibrium is
a possible method for modelling 3D atmospheres in energy equilibrium.

In a series of papers presenting a 3D radiative transfer framework \citep{3d1,3d2,3d3,3d4,3d5,3d6,3d7,3d8},
several radiative-transfer test problems for several scenarios were considered, always under the assumption of 
time independence. In this paper, we extend our framework to include direct time dependence in the solution of radiative transfer
by considering the $\partial I/\partial t$ terms in the 3D radiative transfer equation. 
We used our 1D time-dependent radiative transfer code \citep{jack09} to verify the implementation of time dependence
in the 3D RT framework using a number of test calculations that are discussed below.

In the following section, we describe the method we use to solve the time-dependent 3D radiative transfer.
The test calculations are presented in section 3 to verify that the implementation functions correctly.
In the tests, we include scattering and investigate atmospheres with time-dependent inner boundary conditions.

\section{Transfer equation}

\citet{chen07} present an approach to solve the radiative transfer equation in a flat space time and in the comoving frame.
The radiative transfer equation written in terms of an affine parameter $\xi$, Eq. (18) from \cite{chen07}, is given by
\begin{equation}
\frac{\partial I_{\lambda}}{\partial\xi}+\left(\frac{\dx\lambda}{\dx\xi}\right)\frac{\partial I_{\lambda}}{\partial\lambda}
=-\left(\chi_{\lambda}\frac{h}{\lambda}+\frac{5}{\lambda}\frac{\dx\lambda}{\dx\xi}\right)I_{\lambda}+\eta_{\lambda}\frac{h}{\lambda}.
\end{equation}
The description of the $\partial I/\partial \lambda$-discretization for homologous velocity fields has been presented in \citet{3d5}. We
use this discretization method and extend it to include the $\partial I/\partial t$ time dependence in the solution of the radiative transfer. 

The time-dependent 3D radiative transfer equation along a characteristic is a modification of Eq. (15) in \citet{3d5} and given by
\begin{equation}
\frac{\partial I_{\lambda,t}}{\partial s}+a(s)\frac{\partial}{\partial \lambda}(\lambda I_{\lambda,t})
+a(t)\frac{\partial}{\partial t} I_{\lambda,t}+4a(s)I_{\lambda,t}=-\chi_{\lambda}f(s)I_{\lambda,t}+\eta_{\lambda}f(s),
\end{equation}
where the factor $a(t)$ is simply
\begin{equation}
a(t)=\frac{1}{c}.
\end{equation}
The path length along a characteristic is represented by $s$, the intensity $I_{\lambda,t}$ is a function of the wavelength $\lambda$ and
time $t$ along the characteristic.  For the detailed derivation of $f(s)$ and $a(s)$ see \cite{3d5}.

The fully implicit discretization of equation (1) in both wavelength and time is given by
\[
\frac{\dx I_{\lambda}}{\dx s}+\left[ a(s)\frac{\lambda_{l}}{\lambda_{l}-\lambda_{l-1}}+\frac{a(t)}{\Delta t}+4a(s)+\chi_{\lambda}f(s)\right] I_{\lambda ,t}
\]
\begin{equation}
=a(s)\frac{\lambda_{l-1}I_{\lambda_{l-1}}}{\lambda_{l}-\lambda_{l-1}}+\eta_{\lambda}f(s).
%\label{eq:finite}
\end{equation}
Following \citet{3d5},
this leads to a modification of the effective optical length $\hat{\chi}$, which is now defined by
\begin{equation}
\dx \tau = -\left( \chi_{\lambda}f(s)+4a(s)+\frac{a(s)\lambda_{l}}{\lambda_{l}-\lambda_{l-1}}+\frac{a(t)}{\Delta t}\right)\dx s
\equiv -\hat{\chi}\dx s.
\end{equation}
\citet{3d5} chose an unusual sign convention for $\hat{\chi}$,
  which we have restored to its conventional choice.
The modified source function $\hat{S}_{\lambda}$ is then defined by
\begin{equation}
\frac{\dx I_{\lambda}}{\dx\tau}=I_{\lambda}+\frac{\chi_{\lambda}}{\hat\chi_{\lambda}}\left(S_{\lambda}f(s)+\frac{a(s)}{\chi_{\lambda}}
\frac{\lambda_{l-1}I_{\lambda_{l-1}}}{\lambda_{l}-\lambda_{l-1}}+\frac{a(t)}{\chi_{\lambda}}\frac{1}{\Delta t}I_{t-1}\right)
\end{equation}
\begin{equation}
\equiv I_{\lambda}-\hat{S}_{\lambda}. \label{eqn:rte}
\end{equation}
Our restoration of the conventional sign choice for $\hat{\chi}$
  alters the sign of Eq.~\ref{eqn:rte} as compared to Eq.~21
  of \citet{3d5}.

This approach of including time dependence in the 3D RT framework is similar to the first discretization method for the 1D
case as described in \citet{jack09}. The discretization of the $\partial I/\partial t$ term modifies the generalized optical
depth and adds an additional term to the generalized source function.
Please note that the underlying frame differs in the 1D and 3D
cases. In the 1D case all momentum quantities are measured by a
comoving observer, whereas in the 3D case, only the wavelength is
measured by a comoving observer, the momentum angles are measured in
the observer's frame.
Additionally as noted in \citet{3d5}, the path length differs
in the 3D case and the 1D case. In the 1D case the path length
$ds_M$ is that defined by \citet{mih80}, which is not a true
distance, whereas our $ds$ is a true distance \citep{chen07}.
These differences lead to the differences in the 
equations and in the factor $a(t)$.

\section{Test of the implementation}

The new extension for time dependence in the solution of the 3D radiative transfer needs to be tested
and compared to the results of our time-dependent, 1D spherically
symmetric radiative transfer code.
We implemented the extension as described above into our MPI parallelized Fortran 95 code described in the previous papers
of this series.
For all our test calculations, we use a sphere with a gray continuum opacity
parameterized by a power law in the continuum optical depth $\tau_{std}$.
We interpolate the $\tau_{std}$ profile from the 1D grid onto the 3D
grid using two-point power-law interpolation.  The opacity on the 3D
grid is then given by $\chi=-\Delta \tau/\Delta r$. 
We use a 3D spherical coordinate system because the results can be directly compared
to our 1D spherically symmetric radiation transport code \citep{phhs392}.
The basic model parameters are
\begin{enumerate}
\item Inner radius of $R_{c}=5 \times 10^{10}$~cm and an outer radius of $R_{out}=1 \times 10^{11}$~cm,
\item Optical depth in the range of $\tau_{min}=10^{-6}$ to $\tau_{max}=5$,
\item Grey temperature structure with $T_{model}=10^{4}$ K,
\item Outer boundary condition $I_{bc}^{-}=0$ and diffusive inner boundary condition.
\item We assume an atmosphere in LTE,
\item The atmosphere is static.
\end{enumerate}

For our test calculations, we use a moderately sized 3D grid with
$n_{r}=2*128+1$, $n_{\phi}=2*8+1$ and $n_{\theta}= 2*4+1$ points along each axis,
for a total of $257*9*17\approx2 \times 10^{4}$ voxels.
See section \ref{sec:pert} for a detailed explanation why this voxel setup is used.
In all tests, we use the full characteristic method for the 3D time-dependent RT solution \citep{3d1}.
For the solid angle sampling ($\theta$,\ $\phi$), we chose a grid of $64^{2}$ points, which
is a reasonable resolution of the spherical coordinate system \citep{3d4}.
The time-dependent test calculations are performed with a time step of $10^{-2}\,$s, unless stated
otherwise.

\subsection{Constant atmosphere}

First, we verify the results for the case of a model atmosphere completely
constant in time. The results of the time-dependent RT solution should relax in
time to be equal to the time-independent solution.

% new graph
%
\begin{figure}
\begin{center}
 \resizebox{0.9\hsize}{!}{\includegraphics{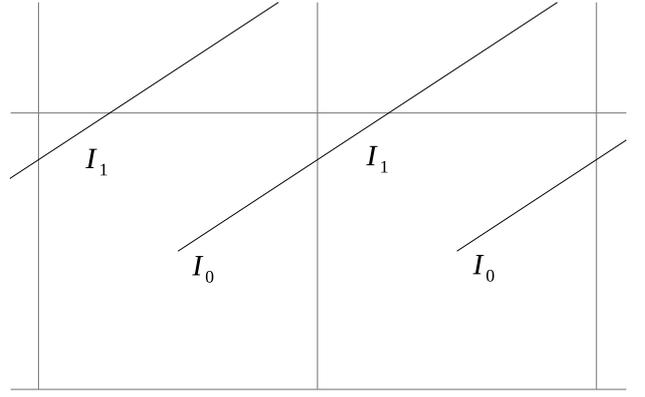}}
\caption{A sketch of two inner voxels and the characteristics of one
  particular angle, $I_0$ and $I_1$ represent
the resulting specific intensities of two characteristics that hit the same voxel.}
\label{fig:char}
\end{center}
\end{figure}

To obtain a numerically accurate solution, it is necessary to follow the
time derivative of the intensities for each characteristic individually (hereafter: subvoxel method).
Previously, the mean intensity of each voxel is a voxel average of the intensities of
all characteristics that go through the particular voxel.
The characteristics are started either from the innermost or outermost voxels.
One specific inner voxel can be hit by many characteristics
of the same angle.
Figure \ref{fig:char} shows an example of two innermost voxels with the characteristics of
one particular angle.
Clearly, the voxels are hit by the characteristic that started at the neighboring voxel.
This second hit gives an additional and different result for specific intensity for the particular voxel.
The intensities are taken at a point of the characteristic that is closest to the center of the voxel.
Previously, the mean of all of these intensities has been used to compute the intensity
of this particular voxel, $I_{\mathrm{mean}}=\sum I_{i}/n$.
Using the average of the intensities passing through a given voxel
to model the time-dependent intensities
results in significant
numerical errors, and must therefore be avoided.
The explanation is that the difference between the averaged intensity and the individual
intensity of a characteristic adds an additional term to the source function. This is completely analogous to the
correct treatment of the wavelength derivative in Lagrange frame radiative transfer
calculations.
In the subvoxel method, all the intensities are
saved and used for the individual characteristics.
The subvoxel method increases the storage requirements, but this is addressed
with a solid-angle domain decomposition method so that the storage requirements
per process remain limited.

\begin{figure}
 \resizebox{\hsize}{!}{\includegraphics{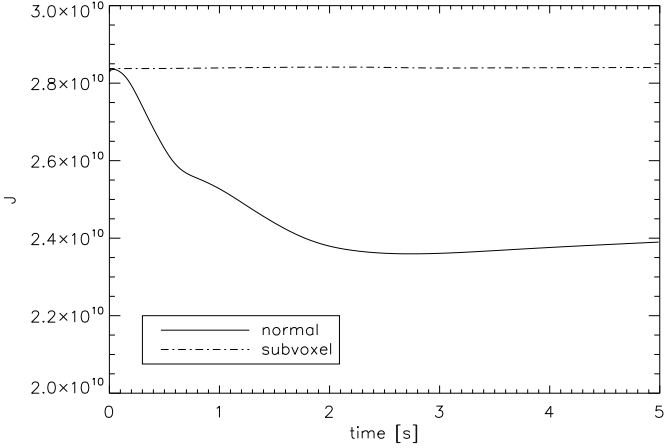}}
\caption{Demonstration of numerical errors. The solid line shows the mean
intensity of one outermost voxel for a method that tracks the intensity time
derivatives by averaging over a voxel. The mean intensity is inaccurate and then
relaxes to the wrong solution.
The dot-dashed line shows the
results of the subvoxel method, which is numerically much more accurate.}
\label{fig:3d_const}
\end{figure}

Figure \ref{fig:3d_const} shows the mean intensity $J$ of one outermost voxel
as function of time. Although the atmosphere structure is time independent,
the mean intensity changes in time and relaxes to a wrong solution
if a numerical method with voxel-averaged
time derivatives is used.
The errors are on the order of 15\% compared to the time independent solution.
The mean intensity obtained with the subvoxel method is
constant in time and identical to the 3D time-independent solution. Thus, we
only use the subvoxel method.

\subsection{Perturbations}
\label{sec:pert}

To see direct effects of the time dependence in the solution of 3D RT problems,
we use a setup with a time variable inner boundary condition of our test model
atmosphere. These perturbations will then propagate through the model
atmosphere by radiation transport in time and finally emerge at the surface of
the sphere. We perform the same calculations with our 1D spherically symmetric
code and compare the results to the time-dependent 3D spherically symmetric RT.

For all our perturbation tests, we place a ``light bulb'' at the center of the model atmosphere.
This light bulb is just a central source of light inside of the atmosphere.
For the first test, the light bulb is switched on and set to (arbitrarily) radiate at $10^{5}$ times the initial
inner boundary intensity. The calculations then track the changes in the 3D (or 1D) radiation field
in time.

\begin{figure}
 \resizebox{\hsize}{!}{\includegraphics{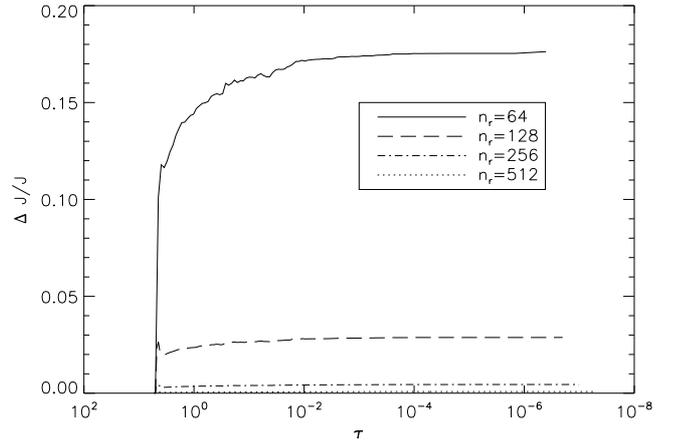}}
\caption{The error of the mean intensity versus $\tau$ for different numbers of radial voxels, $n_{r}$,
of a model atmosphere with a light bulb at the inner boundary that has been
switched on at the start of the calculations.}
\label{fig:3d_nr}
\end{figure}

First, we compare the resulting mean intensity $J$ of the time-independent solution to the relaxed time-dependent
solution, which should be identical.
In Fig. \ref{fig:3d_nr}, the error in the resulting mean intensity, $(J_{dep}-J_{indep})/J_{indep}$,
is shown after the atmosphere has relaxed to the new inner boundary condition.
Clearly the error is lower when a higher resolution in the radial voxels $n_{r}$ is used.
This is equivalent to a higher resolution in the optical depth $\tau$.
With $n_{r}=2*512+1$, the error is reduced to below $0.1\%$.
Therefore, we use a large number of voxels for the radial coordinate. Changing the number of voxels for
the other coordinates does not affect the resulting mean intensity in this spherically symmetric test.

The test model run with $n_{r}=2*512+1$ has been computed on a
supercomputer, where we used 2,048 CPUs.
The computation time for this calculation with 500 time steps is about
two hours. 
The storage requirement is about 11 GB and is mainly required for
saving all the intensities at a time step. 
By using a solid-angle domain decomposition, the  memory per
process required is kept small. 
For true 3D models the resolution in both the other coordinates
$n_{\theta}$ and $n_{\phi}$ likely also needs to be 
significantly higher. These calculations are beyond the
  scope of this work.
In future work, we will test our code with a higher resolution in $n_{\theta}$ and $n_{\phi}$ and
apply realistic 3D models to further test our time-dependent extension.

\begin{figure}
 \resizebox{\hsize}{!}{\includegraphics{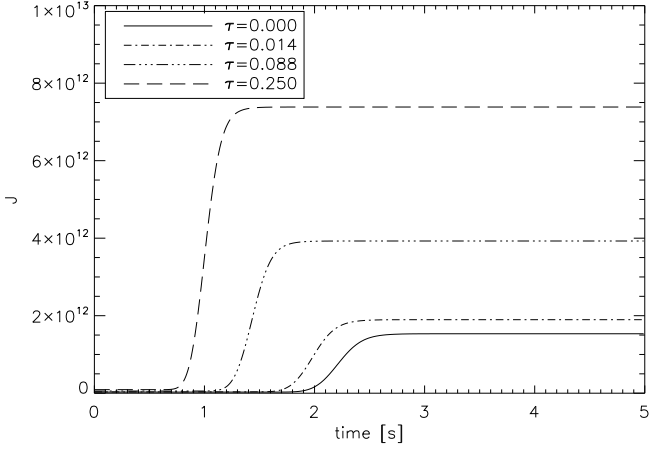}}
\caption{The mean intensity versus time for voxels at different
  optical depth, $\tau$,
of a model atmosphere with a light bulb at the inner boundary that has been
switched on at the start of the calculations.}
\label{fig:3d_bulb}
\end{figure}

\begin{figure}
 \resizebox{\hsize}{!}{\includegraphics{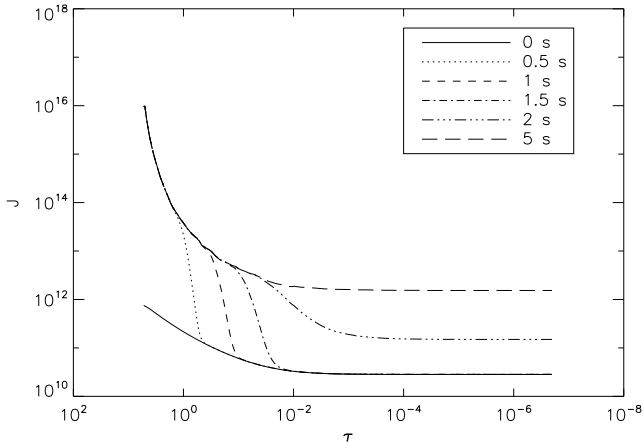}}
\caption{Snapshots of the mean intensity of the atmosphere at
  different points instants in time.} 
\label{fig:3d_bulb_time}
\end{figure}

In Fig. \ref{fig:3d_bulb}, the mean intensity at different layers (radii in the 3D atmosphere) is shown versus time.
The additional radiation of the inner light bulb needs time to propagate through the atmosphere. 
One simple test is to calculate the time the radiation would need at the speed of light to move from
the inner to the outermost layer. For the distance of $R_{out}-R_{c}\approx 5\cdot 10^{10}$~cm, 
the radiation needs about $\approx 1.7\,$s,
which is consistent with the result of the 3D radiative transfer as seen in Fig. \ref{fig:3d_bulb}.
Figure \ref{fig:3d_bulb_time} illustrates the propagation of the radiation ``wave'' throughout the atmosphere.
Here, the mean intensity is plotted against radius for a few snapshots at different times.

\begin{figure}
 \resizebox{\hsize}{!}{\includegraphics{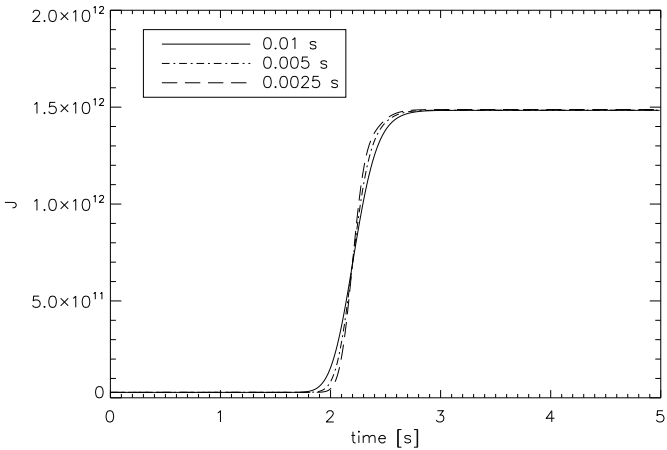}}
\caption{Mean intensity of the outermost radius versus time computed
  with three different time step choices.}
\label{fig:3d_bulb_timestep}
\end{figure}

Another test is to check if the time relaxed result of the time-dependent 3D RT depends
on the size of the time steps.
The results of this test are shown in Fig. \ref{fig:3d_bulb_timestep}, the resulting mean intensities of the outermost layer are plotted
for the test case with the light bulb inside computed
with three different time step sizes.
The change in the mean intensity emerges at the same point in time for all sizes of the time step.
However, the shape of the change is different.
The explanation is that with a shorter time step, the step function
that moves through the atmosphere is 
better resolved in time.

\begin{figure}
 \resizebox{\hsize}{!}{\includegraphics{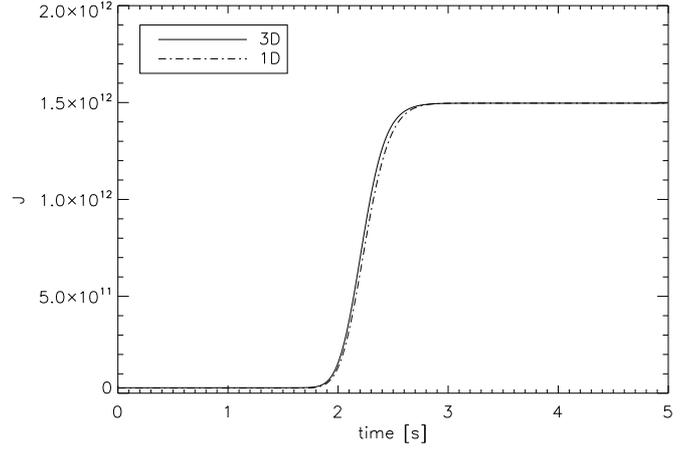}}
\caption{The mean intensity of the outermost radius versus time computed with 3D and 1D time-dependent radiative transfer.}
\label{fig:3d_bulb_1d}
\end{figure}
Another important check is to compare the results of the time-dependent 3D radiative transfer to the
results of our 1D spherically symmetric radiative transfer results.
The mean intensity of the outermost layer of the 3D and the 1D RT time-dependent calculation is shown in Fig. \ref{fig:3d_bulb_1d}.
The test case simulates a light bulb at the inner boundary that has been switched on.
The change in the mean intensity emerges at the same point in time,
and the final mean intensity is the same.

The next interesting test case is to place a sinusoidally varying light bulb at the center of the sphere
and to  follow the radiation field in time.
This leads to an atmosphere where the mean intensity should vary sinusoidally everywhere.
For this test we used a sine wave with a full period of $4$~s.
\begin{figure}
 \resizebox{\hsize}{!}{\includegraphics{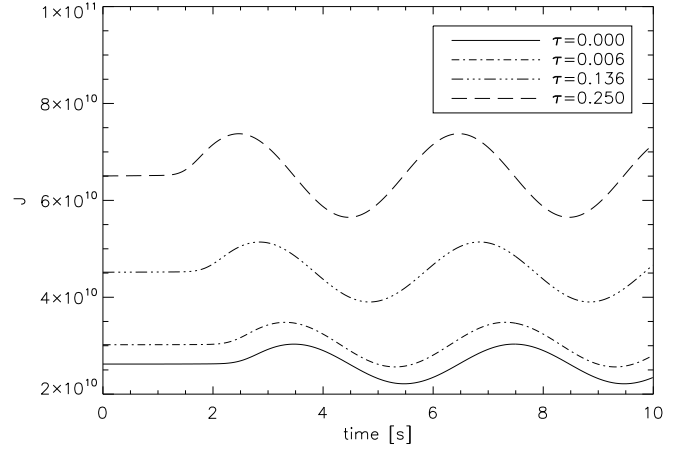}}
\caption{The mean intensity at different optical depths, $\tau$, with a sinusoidally varying light bulb at the center of the sphere.}
\label{fig:3d_sinus}
\end{figure}
\begin{figure}
 \resizebox{\hsize}{!}{\includegraphics{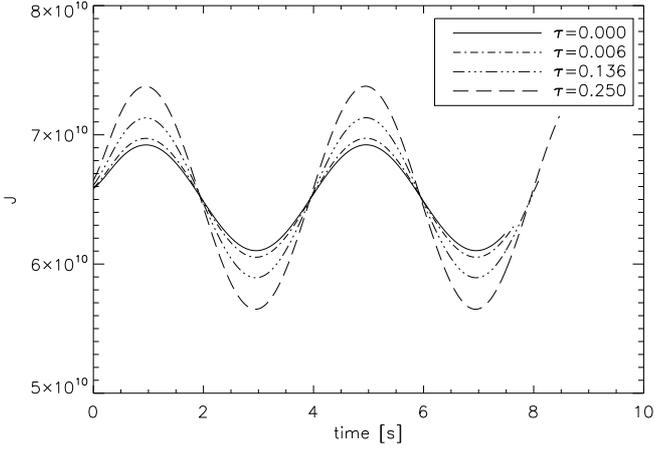}}
\caption{The mean intensity at different optical depths, $\tau$, corrected for the radial time delay. The mean intensities
are shifted arbitrarily to better illustrate the smoothing.}
\label{fig:3d_sinus_cor}
\end{figure}
In Fig. \ref{fig:3d_sinus}, the resulting mean intensities for different radii are  plotted versus time.
It takes about $2$~s before the perturbation has moved outwards to affect the outermost layers.
The sinusoidally varying mean intensity is then observed in every layer.
The phase shift between the inner and outer radii is approximately $\pi$, which is about $2$~s in time.
In Fig. \ref{fig:3d_sinus_cor}, we corrected the resulting sine waves at each optical depth for the radial time delay.
We also shifted the mean intensity arbitrarily to directly overlay the sine waves.
With this correction for the travel time of the light, there is no additional phase shift.
The smoothing of the sine wave as it moves through the atmosphere is also clearly illustrated.
\begin{figure}
 \resizebox{\hsize}{!}{\includegraphics{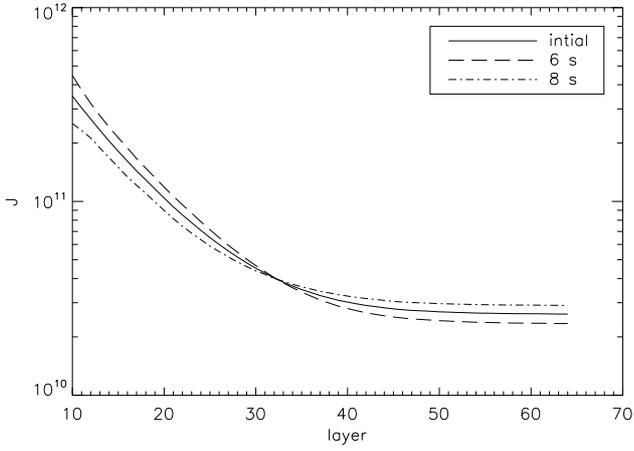}}
\caption{Snapshot of the mean intensity at different phases of the sine wave.}
\label{fig:3d_sinus_time}
\end{figure}
A snapshot of the mean intensity at different moments in time is shown in Fig. \ref{fig:3d_sinus_time}.
The phase difference between the two snapshots is $\pi$.

\subsection{Continuum scattering}

In this section, we investigate the effects of continuum scattering on the solution of the 3D time dependent radiative transfer.
For that, we use a model atmosphere with a low optical depth, so that we can more easily ``see'' the light bulb at the center.
Therefore, we chose for this test: $\tau_{min}=10^{-2}$ and $\tau_{max}=5$.

\begin{figure}
 \resizebox{\hsize}{!}{\includegraphics{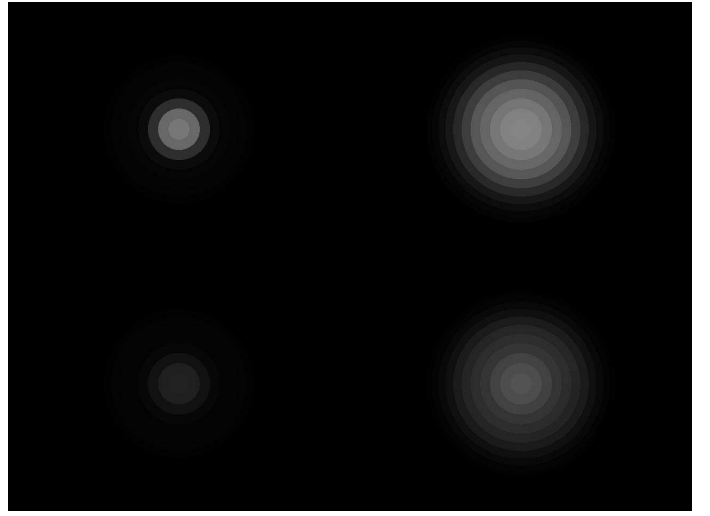}}
\caption{Snapshots of the sphere looking down the pole of the coordinate system. In the right panel side no scattering is considered, while 
in the left panel the results with scattering are shown. The upper row shows a snapshot at time $t=0$~s,
and the lower row is at the time of minimum of the sine wave.}
\label{fig:sinus_both}
\end{figure}
We solved the radiative transfer problem for this model atmosphere both without scattering and for
a scattering dominated atmosphere with $\epsilon=10^{-4}$.
A visualization of both spheres is shown in Fig. \ref{fig:sinus_both}. It shows the image for an observer 
looking down one pole of the coordinate system. The lefthand panel
displays the solution for an atmosphere where no scattering is
considered.
The outer voxels are dark and not visible to the observer, and the sphere shows strong limb darkening. The righthand panel shows the results
with scattering included in the solution of the radiative transfer.
The outer voxels of the disk are significantly brighter as the radiation from the light bulb is scattered towards the observer, thus the model
showing less limb darkening.

\begin{figure}
 \resizebox{\hsize}{!}{\includegraphics{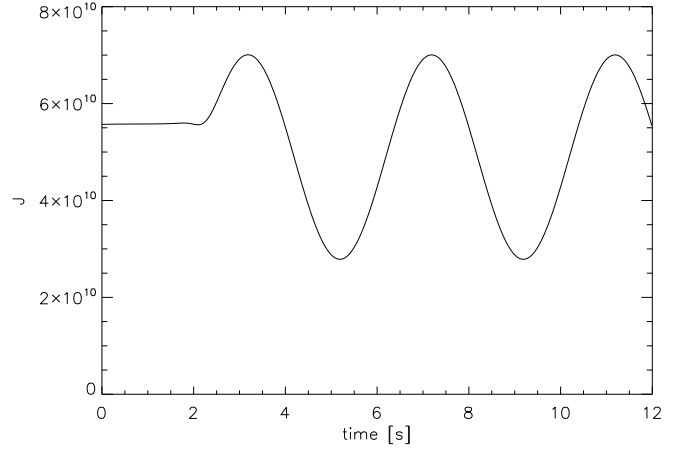}}
\caption{The mean intensity at one outer voxel as a function of time
  for a sinusoidally varying light bulb at the center of the sphere.} 
\label{fig:3d_sinus_noscatter}
\end{figure}
\begin{figure}
 \resizebox{\hsize}{!}{\includegraphics{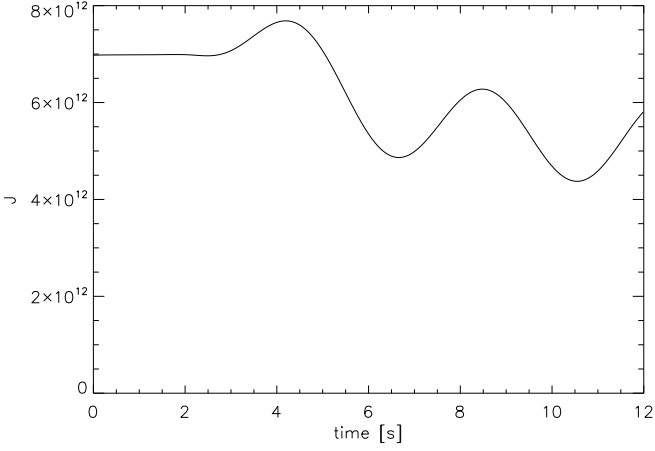}}
\caption{Illustration of the time-dependent mean intensity at the outer boundary for a model with a sinusoidally varying light
bulb at the center. In this solution,
scattering is considered with $\epsilon=10^{-4}$.}
\label{fig:3d_sinus_scatter}
\end{figure}
We now let the light bulb inside of the sphere vary sinusoidally. The resulting mean intensity versus time is 
shown in Fig. \ref{fig:3d_sinus_noscatter} for one outermost
voxel. The mean intensity is varying sinusoidally, as expected. It takes about $2\,$s for the
radiation to travel from the light bulb to the surface.
The resulting mean intensity for the solution of radiative transfer considering scattering with $\epsilon=10^{-4}$
is shown in Fig. \ref{fig:3d_sinus_scatter}.
Again a sinusoidal variation in intensity is seen at the surface, but it has a smaller amplitude than without scattering.
In the presence of scattering, it also takes more time for the radiation to
move through the atmosphere. 
A visualization is shown at Fig. \ref{fig:sinus_both}. The lower row shows the apparent disk at a minimum of the sine.
For the atmosphere where scattering is included, the intensity of the voxels at the outer disk also varies in time.

\subsection{A 3D test model}

All comparisons with our well-tested 1D spherically symmetric
  code show that the results of the test models for our implemented
  time-dependent 3D radiative transfer framework are in good
  agreement. We now want to verify that our 3D time-dependent extension
  also works for test cases of a fully 3D test model
  atmosphere. Unfortunately, we cannot compare these results to our
  1D code, but we can discuss whether the results are reasonable.

  We again use a spherically symmetric setup in the opacity and
  radius for the test model atmosphere.
 For the optical depth, we use a range of $\tau_{min}=10^{-6}$ to $\tau_{max}=5$.
 For the inner radius we chose $R_{c}=5 \times 10^{10}$~cm and for the outer radius $R_{out}=1\;\times\;10^{11}$~cm.
  In this 3D test model, no scattering is considered ($\epsilon=1$).

  Since we now want to examine
  3D effects, we had to increase our angular
  resolution.  For the following test calculation, we used a setup of
  $n_{r}=2*64+1$, $n_{\theta}=2*32+1$, and $n_{\phi}=2*64+1$ voxels.  We
  now place a Gaussian perturbation of the temperature and, therefore,
  the local source function at an off-center position in the sphere.
  The center of the Gaussian has a value for the source function $50$
  times higher than the value of the source function at the center of the sphere
  and is located $20$ voxels away from the center at a radius of $5.69 \times 10^{10}$cm.
  The source function is, therefore, given by
\begin{equation}
S=S(r)+50\cdot S(r)\cdot \exp \{-[(x-x_0)^2+(y-y_0)^2+(z-z_0)^2]/40\},
\end{equation}
with a width for the Gaussian perturbation of $\sigma^2=40$.
 For the first time step, the time-independent model without the perturbation is solved.
 After the first time step, the perturbation
 is introduced and remains constant for the rest of the calculation.
 This means that the resulting mean intensities $J$
 will be constant in time after the relaxation process.
\begin{figure}
 \resizebox{\hsize}{!}{\includegraphics{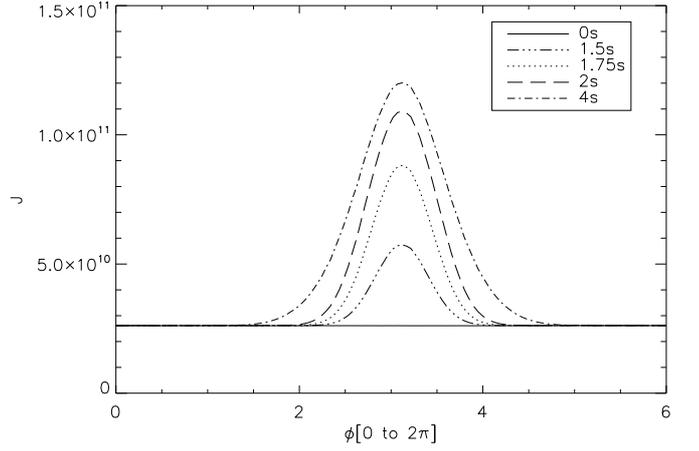}}
\caption{The mean intensity, $J$, at a ring of outermost voxels ($\theta=0$) at different times for a test case with an off-center perturbation.}
\label{fig:3d_3d_test}
\end{figure}
 In Fig. \ref{fig:3d_3d_test}, the mean intensity $J$ of one
  ring of the outermost voxels around the sphere is shown for different
  times. The perturbation is located inside the sphere at
  $\phi=\pi$ and $\theta=0$.
  The mean intensity $J$ at the initial time $t=0s$ is also
 the result for the non-perturbed model.
  As expected, the voxels closer to the perturbation have
  a higher mean intensity $J$. The intensity then decreases, the
  farther away the voxel is from the perturbation. On the other side
  of the sphere, there is no effect on the mean intensities of the
  outer voxels. It is also clear that it takes more time for the
  radiation to reach voxels that are farther away from the
  perturbation.

   The calculation for this  3D test model atmosphere
  run needed about 7,000 CPUh.  This shows how demanding a fully 3D
  time-dependent atmosphere computation is, especially for more
  realistic models in the future.  So far, we have only been able to
 test our time-dependent extension with a gray atmosphere. For realistic
  models, we need to solve the non-gray radiative transfer equation.
  For a type Ia supernova model atmosphere in LTE,
  on the  order of 10,000 wavelength points are needed in the 
  computations. The computation time scales roughly linearly with the
  number of wavelength points.

\section{Conclusion}

We have implemented direct time dependence into our
3D radiative transfer framework.
For best numerical accuracy we used a subvoxel method for the discretized time derivative.
We have also shown that it is important to use  high resolution in the
radial direction. 
To verify the code, we have calculated the solution for
a number of simple parameterized 1D test cases.
The inner boundary condition was made time dependent to simulate radiation waves traveling through an atmosphere.
All these perturbations move through the atmosphere as expected. We also computed a test case with
a sinusoidally varying inner light bulbs.
We compared the results of the 3D time-dependent radiative transfer to the results of our well tested 1D
spherically symmetric radiative transfer program and found excellent agreement.
In a scattering dominated atmosphere, it takes more time for the radiation to move through an atmosphere.
We also calculated a fully 3D test case.
All tests indicate that the new implementations work as intended.

\begin{acknowledgements}
This work was supported in part by the
Deutsche Forschungsgemeinschaft (DFG) via the SFB 676, NSF grant AST-0707704,
and US DOE Grant DE-FG02-07ER41517.
This research used resources of the National Energy Research
Scientific Computing Center (NERSC), which is supported by the Office
of Science of the U.S. Department of Energy under Contract
No. DE-AC02-05CH11231. The computations were carried out at the H\"ochstleistungs Rechenzentrum Nord
(HLRN). We thank all these institutions for generous allocations of
computer time.
We also
thank the anonymous referee for improving the presentation of this work.
\end{acknowledgements}

\bibliographystyle{aa}
\bibliography{18152bib}

\end{document}